# Quantum Imaging with Incoherently Scattered Light from a Free-Electron Laser

Raimund Schneider[1,2], Thomas Mehringer[1,2], Giuseppe Mercurio[3,4], Lukas Wenthaus[3,4], Anton Classen[1,2], Günter Brenner[5], Oleg Gorobtsov[5], Adrian Benz[3], Daniel Bhatti[1,2], Lars Bocklage[5,6], Birgit Fischer[7], Sergey Lazarev[5,8], Yuri Obukhov[9], Kai Schlage[5], Petr Skopintsev[5], Jochen Wagner[10], Felix Waldmann[1], Svenja Willing[5], Ivan Zaluzhnyy[5,11], Wilfried Wurth[3,4,5], Ivan A. Vartanyants[5,11], Ralf Röhlsberger[5,6] & Joachim von Zanthier[*,1,2]

[1]Institut für Optik, Information und Photonik, Universität Erlangen-Nürnberg, Staudtstr. 1, 91058 Erlangen, Germany

[2]Erlangen Graduate School in Advanced Optical Technologies (SAOT), Universität Erlangen-Nürnberg, Paul-Gordan-Straße 6, 91052 Erlangen, Germany

[3]Department Physik, Universität Hamburg, Luruper Chaussee 149, 22761 Hamburg, Germany

[4]Center for Free-Electron Laser Science, Luruper Chaussee 149, 22761 Hamburg, Germany

[5]Deutsches Elektronen-Synchrotron DESY, Notkestr. 85, 22607 Hamburg, Germany

[6]The Hamburg Centre for Ultrafast Imaging, Luruper Chaussee 149, 22761 Hamburg, Germany

[7]Institut für Physikalische Chemie, Universität Hamburg, Grindelallee 117, 20146 Hamburg, Germany

[8]National Research Tomsk Polytechnic University (TPU), pr. Lenina 30, 634050 Tomsk, Russia

[9]Nuclear Safety Institute, Russian Academy of Sciences, B. Tulskaya str. 52, 115191 Moscow, Russia

[10]Institut für Nanostruktur- und Festkörperphysik, Universität Hamburg, Jungiusstr.11, 20355 Hamburg, Germany

[11]National Research Nuclear University MEPhI (Moscow Engineering Physics Institute), Kashirskoe shosse 31, Moscow 115409, Russia

**The advent of accelerator-driven free-electron lasers (FEL) has opened new avenues for high-resolution structure determination via diffraction methods that go far beyond conventional x-ray crystallography methods[1-10]. These techniques rely on coherent scattering processes that require the maintenance of first-order coherence of the radiation field throughout the imaging procedure. Here we show that higher-order degrees of coherence, displayed in the intensity correlations of incoherently scattered x-rays from an FEL, can be used to image two-dimensional objects with a spatial resolution close to or even below the Abbe limit. This constitutes a new approach towards structure determination based on incoherent processes[11,12], including Compton scattering, fluorescence emission or wavefront distortions, generally considered detrimental for imaging applications. Our method is an extension of the landmark intensity correlation measurements of Hanbury Brown and Twiss[13] to higher than second-order paving the way towards determination of structure and dynamics of matter in regimes where coherent imaging methods have intrinsic limitations[14].**

The discovery by Hanbury Brown and Twiss of photon bunching of thermal light[15] and its use for the determination of the angular diameter of stars by measuring spatial photon correlations[13] was a hallmark experiment for the development of modern quantum optics[16]. The subsequent quantum mechanical description of photon correlations by Glauber paved the way for a generalised concept of optical coherence[17] that is founded on the analysis of correlation functions of order $m$ rather than the 1$^{st}$-order coherence. For example, the spatial 2$^{nd}$-order photon correlation function $g^{(2)}(r_1, r_2)$ expresses the probability to detect a photon at position $r_1$ given that a photon is recorded at position $r_2$. In the case of two incoherent sources $g^{(2)}(r_1, r_2)$ displays a cosine modulation which oscillates at a spatial frequency depending on the source separation[18,19]. In this way interference fringes show up even in the complete absence of 1$^{st}$-order coherence, allowing for the extraction of structural information from incoherently emitting objects. This has been applied in earth-bound stellar interferometry to measure the angular diameter of stars with 100-fold increased resolution[13] or to reveal the spatial and statistical properties of pulsed FEL sources[20,21].

Extending this concept to arbitrary arrangements of incoherently scattering emitters enables one to use intensity correlations for imaging applications. This has been demonstrated recently for one-dimensional arrays of emitters in the visible range of the spectrum[22-24] where a spatial resolution even below the canonical Abbe limit has been achieved. Here, we go still further and employ the method to image arbitrary two-dimensional objects which incoherently scatter XUV radiation, thus opening up a new approach for x-ray structure determination.

The experiment has been performed at the PG2 beamline of the FLASH FEL facility at DESY, Hamburg[25]. The scheme of the setup is shown in Fig. 1a. The FEL runs in a 10 Hz pulsed mode ($\tau_{pulse} \approx$ 60 fs) at a center wavelength of $\lambda$ = 13.2 nm (for experimental details see[21]). After passing a monochromator to limit the bandwidth to 0.1% (± 0.013 nm), the FEL beam impinges on a slowly moving diffusor converting the spatially highly coherent light of the FEL to quasi-monochromatic pseudo-thermal light with a far field speckle pattern that fluctuates randomly from shot to shot but is stationary for each pulse[26]. This ensures that any sets of scatterers illuminated by the light field from the diffusor act as ensembles of incoherent sources emitting independently from each other (see Supplementary material). In our experiment, a two-dimensional object mask is placed behind the diffusor, consisting of six square-cut holes in a hexagonal arrangement to mimic the carbon atoms in a benzene molecule emitting incoherent fluorescence radiation. The structure of this object is characterised by a set of nine spatial frequency vectors $\zeta = \{\boldsymbol{f}_i = (f_{i,x}, f_{i,y}) = (h_{i,x}/\lambda L, h_{i,y}/\lambda L), i = 1, ...,9\}$, determined by the nine different source separation vectors $\boldsymbol{h}_i = (h_{i,x}, h_{i,y})$, $i = 1, ...9$, connecting the artificial atoms, i.e., the holes in the mask, as shown in Fig. 1b.

The resulting intensity patterns are recorded in the far-field for each pulse individually (see Fig. 2a). By correlating the intensities $I(\boldsymbol{r}_i)$ at different positions $\boldsymbol{r}_i$, $i = 1, ..., m$, for each pulse and averaging the result over many pulses we obtain the $m^{th}$-order intensity correlation function

$$g^{(m)}(\boldsymbol{r}_1, ..., \boldsymbol{r}_m) = \frac{\langle I(\boldsymbol{r}_1) \cdots I(\boldsymbol{r}_m) \rangle}{\langle I(\boldsymbol{r}_1) \rangle \cdots \langle I(\boldsymbol{r}_m) \rangle}. \quad (1)$$

Our imaging algorithm is based on extracting the set of spatial frequency vectors $\zeta$ of the unknown source distribution – assumed to be placed on a 2D lattice – from the higher-order intensity correlations $g^{(m)}$; the source geometry in real space is then reconstructed from $\zeta$ (see Supplementary material). In the intensity correlations of $2^{nd}$ order the entire set of spatial frequency vectors appears, hence the number of parameters to be fitted and extracted from $g^{(2)}$ is particularly large, especially in the case of a data set with low signal to noise ratio. In contrast, measuring intensity correlations of higher than 2nd order allows to split the information, as only a subset of spatial frequencies appears within a given correlation order $m$ (see below). This leads to a significantly reduced number of fit parameters to be determined from each correlation function $g^{(m)}$, allowing to reveal $\zeta$ with substantially increased accuracy[24]. In certain cases even a numerical aperture smaller than required by the canonical Abbe limit allows to determine $\zeta$, leading to a spatial resolution below the classical diffraction limit (see Supplementary material).

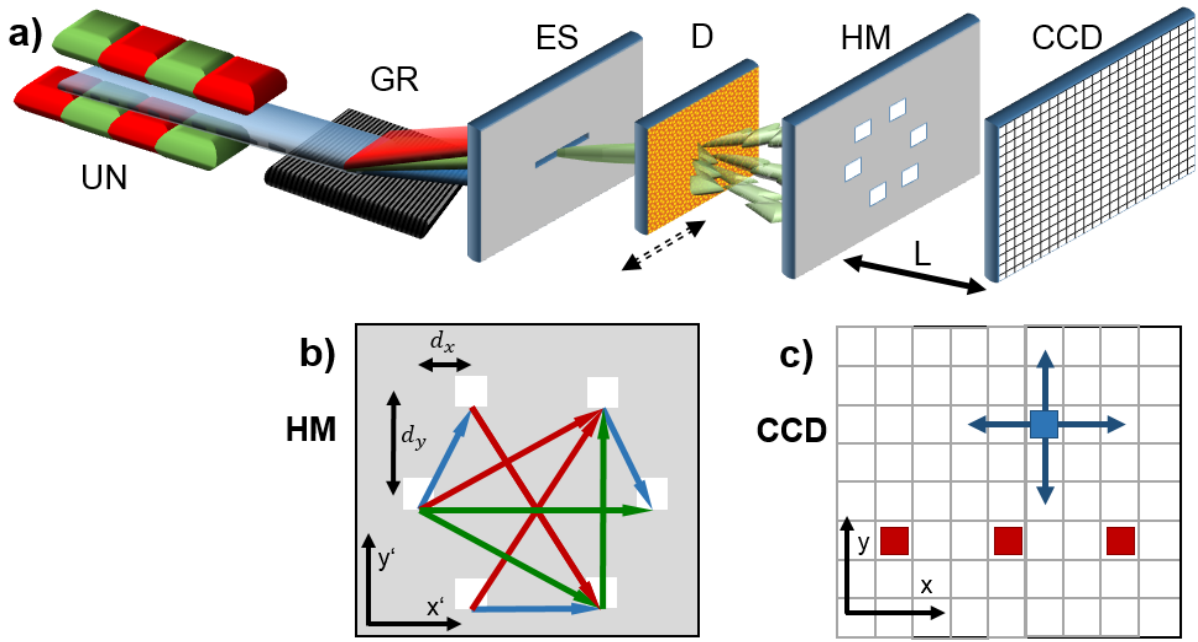

**Figure 1: Scheme of the experiment. a):** The FEL beam created in the undulator (UN) passes a monochromator consisting of a grating (GR) and an exit slit (ES) such that quasi-monochromatic coherent radiation impinges on the moving diffusor (D). The pseudo-thermal light scattered by the diffusor is used to illuminate a benzene-like hole mask (HM) generating six quasi-monochromatic independently radiating incoherent light sources. The resulting light fields are measured at a distance $L$ behind HM by a CCD. **b):** The nine different source separation vectors $\boldsymbol{h}_i$, $i = 1, \ldots 9$, of HM are shown in different colours for a better readability. **c):** To determine $g^{(m)}(\boldsymbol{r}_1; MP_x)$, for each CCD image, $m - 1$ pixels of the CCD camera are selected at the magic positions along the x-axis (red) and correlated with all other pixels (blue) of the CCD; thereafter an average over all images is taken.

The light scattered by the artificial atoms, i.e., the holes in the mask, is measured by a CCD camera in the far field at a distance $L = 275$ mm behind the source plane, where each pixel of the CCD serves as an independent detector. The $m^{th}$-order spatial intensity correlation functions $g^{(m)}(r_1, ..., r_m)$ are computed by correlating the intensities measured by $m$ pixel detectors at positions $r_i$, $i = 1, ..., m$, for each CCD image individually and then averaging over all CCD images (see Supplementary material).

In general $g^{(m)}(r_1, ..., r_m)$ takes a complicated form, depending on the detector positions $r_1, ..., r_m$ and the source geometry. However, when placing all but one detectors at the so-called magic positions (MP)[23,24], only specific spatial frequency vectors of the object appear within a given correlation function of order m. In particular, if $m - 1$ detectors are placed along the x-axis such that $\tilde{f}_x x_j = \frac{j-2}{m-1}$, $j = 2, ..., m$, where $\tilde{f}_x = d_x/(\lambda L)$ is the spatial frequency associated with the lattice constant $d_x$, only those spatial frequency vectors $f_i$ of the object appear in $g^{(m)}(r_1; MP_x)$ where the x-component fulfils the filtering condition $f_{i,x} = \kappa(m-1)\tilde{f}_x$, with $\kappa \in \mathbb{N}_0$; an analogue filtering condition holds for the y-direction. As in 1D[24], the filtering process in 2D is a result of sum rules which hold for the (m-1)$^{th}$-roots of unity (see Supplementary material).

As an example, we consider $g^{(4)}(r_1; MP_x)$ with three detectors located at the MP along the x-direction (see Fig. 1). In this case only the spatial frequency vectors $\zeta_x^{(4)} = \{\binom{3}{1}, \binom{3}{-1}, \binom{0}{2}\}$ (in units of $\tilde{f}_x$ and $\tilde{f}_y$) appear in $g^{(4)}(r_1; MP_x)$, as only $f_x = 1 \cdot 3\tilde{f}_x$ (with $\kappa = 1$) and $f_x = 0 \cdot 3\tilde{f}_x$ (with $\kappa = 0$) match the filtering condition. Note that even though the fixed detectors are aligned at the MP along the x-direction, the corresponding components $f_y$ of the filtered $f_x$ also show up in $g^{(4)}(r_1; MP_x)$.

After evaluating $g^{(m)}(r_1; MP)$ for $m = 3, 4, 5$, with the fixed detectors aligned at the MP along the x-axis as well as along the y-axis (resulting in a total of six 2D correlation functions), the set of spatial frequency vectors $\zeta_{exp}$ of the benzene structure is derived from a best 2D fit to the experimental data (see Supplementary material). As an example, the correlation function $g^{(4)}(r_1; MP_x)$, experimentally determined after evaluating 10.800 single shot CCD images with three pixel detectors located at the MP along the x-direction, and corresponding fit functions are shown in Fig. 2(b-d).

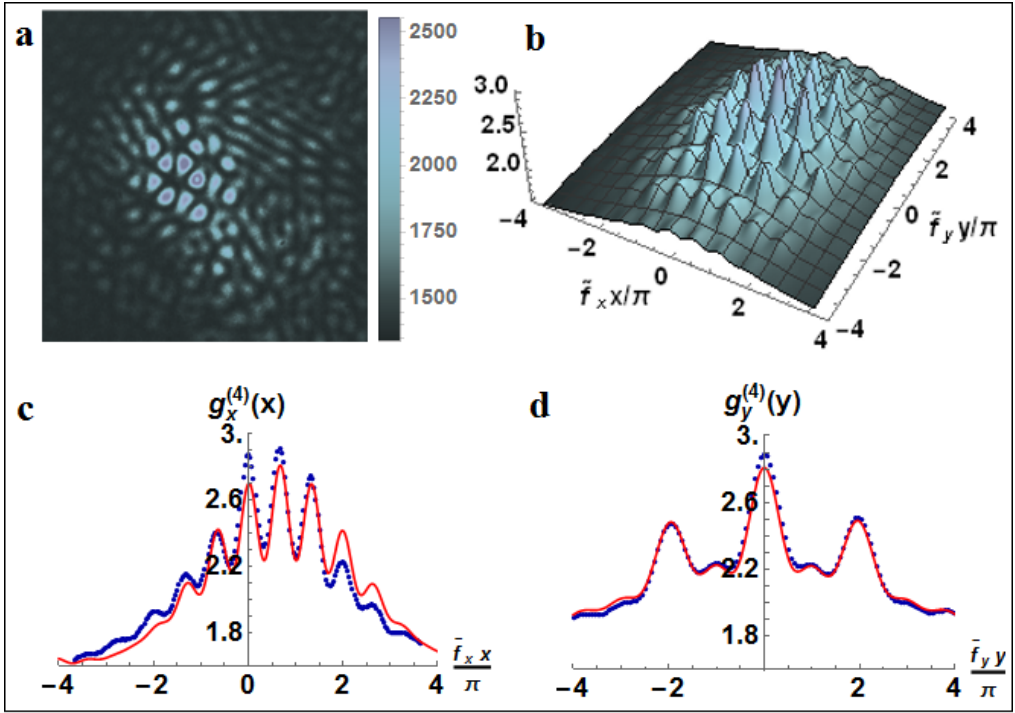

**Figure 2: Single shot speckle pattern and experimentally derived 4$^{th}$-order intensity correlation function $g^{(4)}(r_1; MP_x)$.** **a): S**ingle shot speckle pattern as measured by the CCD. **b):** 10.800 single shot speckle patterns are processed to obtain $g^{(4)}(r_1; MP_x)$ where three pixel detectors are located at the MP along the x-direction. **c) and d):** cross-sections of $g^{(4)}(r_1; MP_x)$ (blue dotted curve) and of the best 2D fit to $g^{(4)}(r_1; MP_x)$ (red solid curve) in x-direction (at $\tilde{f}_y y = 0$) and in y-direction (at $\tilde{f}_x x = \frac{2}{3}\pi$), respectively. The 2D fit allows to access the set of spatial frequencies $\zeta_x^{(4)}$ displayed by $g^{(4)}(r_1; MP_x)$.

The set of spatial frequency vectors $\zeta_{exp}$ thus obtained is shown in Tab. 1. Note that since some spatial frequencies match the filtering condition for more than one correlation order they can be accessed from correlation functions $g^{(m)}(r_1; MP)$ of different orders $m$. However, it turns out, that the values for $f_x$ and $f_y$ thus obtained deviate from each other by less than 1%.

**Table 1: Theoretically expected and experimentally obtained spatial frequency vectors (in units of $\tilde{f}_x$ and $\tilde{f}_y$); colors correspond to the colors of the spatial frequency vectors shown in Fig. 1**

| $\zeta$ | $\binom{1}{1}$ | $\binom{1}{-1}$ | $\binom{2}{0}$ | $\binom{2}{2}$ | $\binom{2}{-2}$ | $\binom{3}{1}$ | $\binom{3}{-1}$ | $\binom{4}{0}$ | $\binom{0}{2}$ |
|---|---|---|---|---|---|---|---|---|---|
| $\zeta_{exp}$ | | | $\binom{1.94}{0.02}$ | $\binom{1.96}{2.00}$ | $\binom{1.97}{-1.95}$ | $\binom{2.97}{0.97}$ | $\binom{2.97}{-1.03}$ | $\binom{3.90}{0.04}$ | $\binom{0.01}{1.98}$ |

A systematical error for the experimentally derived set of spatial frequencies $\zeta_{exp}$ originates from the finite size of the pixel detectors, as this prevents the $m$-1 fixed pixel detectors from being located exactly at the MP (assumed to be point-like). In our experiment, using a CCD camera with pixel size $13.5 \times 13.5\ \mu m^2$ at a distance of $L = 275\ mm$ behind the hole-mask, this results in a systematical error for $f_x$ and $f_y$ of 3.6% and 6.3%, respectively.

Note that according to the filtering condition the two spatial frequency vectors $\binom{1}{1}$ and $\binom{1}{-1}$ (in units of $\tilde{f}_x$ and $\tilde{f}_y$) cannot be extracted from $g^{(m)}(\mathbf{r}_1; MP)$, if $m \geq 3$. Hence, without measuring $g^{(2)}(\mathbf{r}_1; MP)$, four different possibilities to complete the set $\zeta$ from the set $\zeta_{exp}$ are possible, containing either none, one of the two, or both spatial frequency vectors. However, for the investigated benzene structure only the set containing both spatial frequency vectors $\binom{1}{1}$ and $\binom{1}{-1}$ provides a solution for the source arrangement in real space, resulting in a unique set $\zeta$ (see Supplementary material).

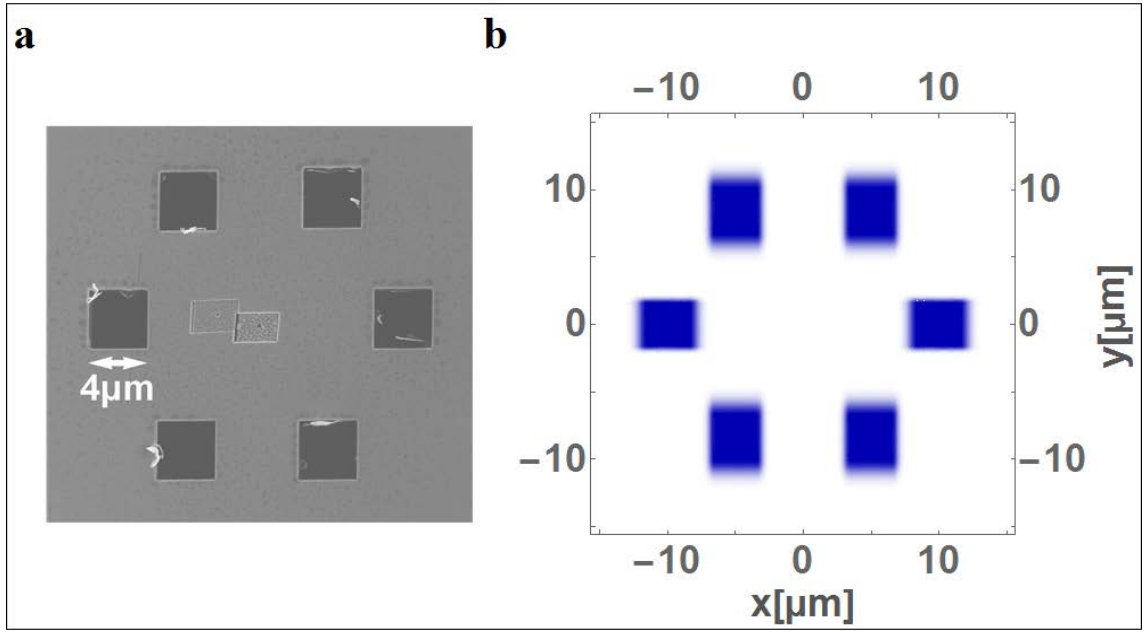

**Figure 3: Geometry of the artificial benzene molecule.** *a)*: Scanning electron microscopy image of the hole mask. *b)*: Image of the hole mask reconstructed after evaluating $g^{(m)}(\mathbf{r}_1; MP)$ for m = 3,4,5, with the fixed detectors aligned at the MP along the x-axis as well as along the y-axis. The uncertainties of the source positions and sizes are displayed by a Gaussian shaped color gradient with a standard deviation equal to the corresponding error (for details see Supplementary material).

Besides the positions of the sources, the correlation signals also contain information about the size of the sources as their finite extension translates into sinc-shaped envelopes in x- and y-direction of $g^{(m)}(\boldsymbol{r}_1; MP)$ (see Supplementary material). Merging the extracted fit parameters, i.e., the spatial frequencies of $g^{(m)}(\boldsymbol{r}_1; MP)$ for m = 3, 4, 5, as well as their envelopes, we are able to determine the entire benzene structure. The result is shown in Fig. 3.

The reconstruction of the source geometry using our approach proves to be extremely robust against intensity and frequency fluctuations of the scattered light[24]. In particular, the method allows to reconstruct the source arrangement in complete absence of 1st-order coherence, e.g., due to imperfect optics that leads to wave front distortions or vanishing coherent diffraction signals. Such aspects become increasingly important with decreasing wavelength. Thus imaging applications in the regime of hard x-rays, where 1st-order coherence is readily compromised by imperfect optical elements, will especially benefit from our imaging technique.

The method has further great potential for imaging with intense sources of x-rays, where a significant part of the emission and scattering occurs incoherently due to inelastic processes like Compton scattering or associated with the strong electromagnetic driving[11]. The approach will therefore become increasingly relevant for imaging with existing and future laser sources in the regime of hard x-rays, especially with respect to the ultimate goal of single-molecule imaging[9,10]. Note that the method can also be combined with novel techniques like macromolecular imaging from imperfect crystals[27] where the single-molecule scattering signal is discriminated with high signal-to-noise ratio against the coherent Bragg peaks from the crystalline structure. Moreover, applying the technique to the detection of resonance fluorescence as an intrinsically incoherent scattering process enables element-specific imaging applications. Finally, it should be emphasised that the role of photons can be taken by any other bosonic particle. Using correlation functions based on anti-commutation relations, our imaging method could be further extended to fermions, i.e., pulsed beams of electrons and possibly also pulsed neutrons from spallation neutron sources.

# References

1. Chapman, H. N. *et al.* Femtosecond diffractive imaging with a soft-X-ray free-electron laser. *Nature Phys.* **2**, 839–843 (2006).
2. Chapman, H. N. *et al.* Femtosecond X-ray protein nanocrystallography. *Nature* **470**, 73–77 (2011).
3. Seibert, M. *et al.* Single mimivirus particles intercepted and imaged with an X-ray laser. *Nature* **470**, 78–81 (2011).
4. Loh, N. D. *et al.* Fractal morphology, imaging and mass spectrometry of single aerosol particles in flight. *Nature* **486**, 513–517 (2012).
5. Kupitz, C. *et al.* Serial time-resolved crystallography of photosystem II using a femtosecond X-ray laser. *Nature* **513**, 261-265 (2014).
6. Takahashi, Y. *et al.* Coherent diffraction imaging analysis of shape-controlled nanoparticles with focused hard X-ray free-electron laser pulses. *Nano Lett.* **13**, 6028–6032 (2013).
7. Barke, I. *et al.* The 3D-architecture of individual free silver nanoparticles captured by X-ray scattering. *Nat. Commun.* **6**, 6187 (2015).
8. Neutze R., Wouts, R., van der Spoel, D., Weckert, E. & Hajdu, H. Potential for biomolecular imaging with femtosecond X-ray pulses. *Nature* **406**, 752-757 (2000).
9. Aquila, A. et al. The linac coherent light source single particle imaging road map. *Struct. Dyn.* **2**, 041701 (2015).
10. Barty, A., Küpper, J. & Chapman, H. N. Molecular Imaging Using X-ray Free-Electron Lasers. *Annu. Rev. Phys. Chem.* **64**, 415–435 (2013).
11. Slovik, J. M., Son, S.-K., Dixit, G., Jurek, Z. & Santra, R. Incoherent x-ray scattering in single molecule imaging. *New J. Phys.* **16**, 073042 (2014).
12. Gorobtsov, O.Y., Lorenz, U., Kabachnik, N. M., Vartanyants, I. A. Theoretical study of electronic damage in single-particle imaging experiments at x-ray free-electron lasers for pulse durations from 0.1 to 10 fs. *Phys. Rev. E* **91**, 062712 (2015).
13. Hanbury Brown, R. & Twiss, R. Q. A Test of a New Type of Stellar Interferometer on Sirius. *Nature* **178**, 1046-1048 (1956).
14. Chapman, H. N., Nugent, K. A. Coherent lensless X-ray imaging. *Nat. Photonics* **4**, 833-839 (2010).
15. Hanbury Brown, R. & Twiss, R. Q. Correlation between photons in two coherent beams of light. *Nature* **177**, 27-29 (1956)
16. Glauber, R. J. Nobel Lecture: One hundred years of light quanta. *Rev. Mod. Phys. 78, 1267-1278* (2006).
17. Glauber, R. J. The Quantum Theory of Optical Coherence. *Phys. Rev.* **130**, 2529 (1963)
18. Goodman, J. W*., Statistical Optics*. John Wiley & Sons, New York (1985).
19. Baym, G. The physics of Hanbury Brown-Twiss intensity interferometry: from stars to nuclear collisions. *Acta Phys. Pol., B* **29**, 1839-1884, (1998).
20. Singer, A. *et al*. Hanbury Brown-Twiss Interferometry at a Free-Electron Laser. *Phys. Rev. Lett.* **111,** 034802 (2013).

21. Gorobtsov, O.Y. *et al.* Statistical properties of a free-electron laser revealed by Hanbury Brown-Twiss interferometry. *Phys. Rev. A* **95**, 023843 (2017).
22. Thiel, C. *et al.* Quantum Imaging with incoherent photons. *Phys. Rev. Lett.* **99**, 133603 (2007)
23. Oppel, S., Büttner, T., Kok, P. & von Zanthier, J. Superresolving Multiphoton Interferences with Independent Light Sources. *Phys. Rev. Lett.* **109**, 233603 (2012).
24. Classen, A. *et al.* Superresolving Imaging of Irregular Arrays of Thermal Light Sources using Multiphoton Interferences. *Phys. Rev. Lett.* **117**, 253601 (2016).
25. Ackermann, W. *et al.* Operation of a free-electron laser from the extreme ultraviolet to the water window. *Nat. Photonics* **1**, 336–342 (2007).
26. Goodman, J.W. *Speckle Phenomena in Optics: Theory and Applications*, Roberts and Company, Englewood, Colorado (2007).
27. Ayyer, K. *et al.* Macromolecular diffractive imaging using imperfect crystals, Nature 530, 202-206 (2016).
28. Pearce, M. E., Mehringer, T., von Zanthier, J. & Kok, P. Precision estimation of source dimensions from higher-order intensity correlations. *Phys. Rev. A* **92**, 043831 (2015).

**Acknowledgements**

R.S., T.M, A.C., D.B. J.v.Z. gratefully acknowledge funding by the Erlangen Graduate School in Advanced Optical Technologies (SAOT) by the German Research Foundation (DFG) in the framework of the German excellence initiative. A.C. and D.B. gratefully acknowledge financial support by the Staedtler Foundation and the Cusanuswerk, Bischöfliche Studienförderung, respectively. We acknowledge support of the Helmholtz Association through project oriented funds. I.V. acknowledge support of the Virtual Institute VH-VI-403 of the Helmholtz Association. Y.O. and S.W. acknowledge support by the Partnership for Innovation, Education and Research (PIER) between DESY and the University of Hamburg. We are grateful to the FLASH machine operators, to the technical staff at FLASH for excellent FEL conditions and to Holger Meyer for his contributions to the design of the experimental setup. We appreciate fruitful discussions with E. Weckert and H. N. Chapman.

# Supplementary material

**Creation of x-ray photons**

To generate XUV photons, the FLASH FEL was operated in a 10 Hz single pulse mode with six undulator modules and total undulator length of 30 m. The electron bunch charge was 0.3 nC, and the electron beam energy 678.0 MeV. The average photon pulse length, deduced from a statistical analysis of the FEL pulses[21], was below 60 fs (FWHM) and the average photon pulse energy 25 µJ. The central wavelength $\lambda$, measured at the PG2 beamline in spectrometer mode, was 13.2 nm (93.93 eV) with a spectral bandwidth of 1.1%. With an exit slit opening of 300 µm, the monochromator settings allow a bandpass filtering to 0.096% (0.09 eV). With four C-coated mirrors, one grating and the slit bandpass used, the beamline transmission was 2%.

**Generation of pseudo-thermal light**

The coherent x-ray beam impinges on a diffusor produced from spin coating a $Si_3N_4$ substrate of size 10 x 10 mm² and 200 µm thickness with a solution of silica ($SiO_2$) nanospheres of diameter 200 nm in ethanol. The photons from the coherent FEL beam were scattered with a random phase from the homogeneous monolayer of silica spheres, resulting in a speckle pattern with a Gaussian field distribution in the source plane as well as in the detection plane (see Figs. 1 and 2(a) of the main text)[18,19,26]. To ensure that each single shot is an independent realisation of a chaotic light field, the diffusor was constantly moved laterally with a speed of 0.02 mm/s during measurements so that every laser pulse impinges on a different part of the diffuser and creates during each laser pulse a different stationary speckle pattern (see Fig. 2(a)).

A mask with six holes in a hexagonal structure mimicking a benzene molecule was placed behind the diffusor. The mask was made by perforating a $Si_3N_4$ substrate of 200 nm thickness with focused ion beam technique producing six free-standing square holes with equal side length $a = 4.0$ µm, arranged in a hexagonal geometry with grid constants $d_x = 5.0$ µm and $d_y = 8.7$ µm in x- and y-direction, respectively (see Fig. 1). The opacity of the mask is increased by evaporating ≈ 340 nm thick gold layer onto the mask's surface.

The mask was located at a distance of 10 mm behind the diffusor ensuring that many coherence cells are located within the area of each hole. In this way, each hole acts as an independent source of pseudo-thermal light, whose lateral coherence length $l_c$ in the far field is defined according to van Cittert-Zernike theorem by the side lengths of the holes[18].

**Measurement of higher-order intensity correlations**

In order to record higher order intensity correlations, a CCD camera (Andor iKon-L 936) with 2048 × 2048 pixels of size 13.5 × 13.5 µ$m^2$ was placed behind the hole mask at a distance $L = 275$ mm. This yields a lateral coherence length of $l_c \approx 1.3$ mm at the CCD which extends over many pixels, a prerequisite to resolve the higher-order intensity correlations. Due to the limited read out speed of the CCD, a dynamic beam shutter with a frequency of 1 Hz was used, blocking the beam for the whole exposure but the duration of a single pulse. Each pixel of the CCD can be considered to be an individual photon detector. In order to calculate the $m^{th}$-order spatial intensity correlation function $g^{(m)}(\boldsymbol{r}_1, \ldots, \boldsymbol{r}_m) = \frac{\langle I(\boldsymbol{r}_1) \cdots I(\boldsymbol{r}_m) \rangle}{\langle I(\boldsymbol{r}_1) \rangle \cdots \langle I(\boldsymbol{r}_m) \rangle}$, the intensities $I(\boldsymbol{r}_i)$ (grey values) measured by $m$ pixels of the CCD at positions $\boldsymbol{r}_i$, $i = 1, \ldots, m$, were multiplied and the products thereafter averaged over the whole data set, consisting of the central area (300 x 300 pixels) of all CCD images (10.800 uncompressed 16-bit greyscale images). To enhance the statistics the concept of spatial averaging was applied. This means that each image of 300 x 300 pixels is divided into 441 partially overlapping sub regions of size 200 x 200 pixels. A preliminary correlation function is calculated by processing only one subregion for the whole set of 10.800 images. This is repeated for all subregions. The final correlation function is then obtained by averaging over all 441 preliminary correlation functions, which is possible since $g^{(m)}$ depends only on relative detector positions. Note that applying this method does not affect the spatial frequency determination used in our imaging algorithm[28]. The normalisation factor appearing in the denominator of $g^{(m)}(\boldsymbol{r}_1, \ldots, \boldsymbol{r}_m)$ was obtained by taking into account the average grey value for each individual detector pixel at position $\boldsymbol{r}_i$, $i = 1, \ldots, m$.

**Imaging algorithm**

To apply our imaging algorithm we need to translate particular pixel positions on the CCD to detector positions. The phases of the fixed detectors at the magic positions (MP) along the x-axis are given by $\tilde{f}_x x_j = \frac{j-2}{m-1}$ for $j = 2, \ldots, m$, where $\tilde{f}_x = d_x/(\lambda L)$ is the spatial frequency associated with the lattice constant $d_x$. To determine $d_x$ experimentally, we investigated the correlation functions of 3$^{rd}$ order along the x-axis. We started with the two fixed detector pixels located approximately at the same position and spread them out in subsequent evaluations of $g^{(3)}(x_1, x_2, x_3)$. We exploit the fact that if $m - 1$ equidistantly separated pixels match the MP all spatial frequencies $f_x$ not fulfilling the filtering condition $f_x = \kappa(m-1)\tilde{f}_x$ are suppressed[24]. Extracting the corresponding pixel positions for $m = 3$ we determine the grid constant via $d_x = \frac{\lambda L}{2 x_3}$ from the position $x_3$ of the third detector. This process can be applied equally to the y-direction, providing access to $d_y$. Knowing $d_x$ and $d_y$, the pixel positions representing the MP for any correlation order $m$ can be calculated in the x- and the y-direction.

In the evaluation of the experiment, the m-1 fixed pixels were set to the MP along the x-axis in the center of a given subregion, and two-dimensional correlation functions $g^{(m)}(\mathbf{r}_1; MP_x)$ were evaluated for $m = 3,4,5$. Employing the filtering condition $f_{i,x} = \kappa(m-1)\tilde{f}_x, \kappa \in \mathbb{N}_0$, this allows to access all spatial frequency vectors $\mathbf{f}_i$ of the benzene molecule of which the x-component $f_{i,x}$ is a multiple of 0, 2, 3, 4. Equally, $g^{(m)}(\mathbf{r}_1; MP_y)$ were determined for $m = 3,4,5$, with m-1 pixels fixed to the MP along the y-axis in the center of a given subregion, providing information about all spatial frequency vectors $\mathbf{f}_i$ of the benzene molecule of which the y-component $f_{i,y}$ is a multiple of 0, 2, 3, 4. To retrieve the spatial frequency vectors $\mathbf{f}_i = (f_{i,x}, f_{i,y})$ of our benzene structure from the measured correlation signals, two-dimensional functions of the form

$$g^{(m)}_{fit}(x,y) = c_0 + env(x,y,w_x,w_y) \times \left[\sum_i A_i \times \cos(\pi f_{i,x} x + \pi f_{i,y} y)\right] + u_0 \times env(x,y,u_x,u_y) \ ;$$

$$env(x,y,w_x,w_y) := sinc\left[\left(\pi \tilde{f}_x x - \frac{m-2}{m-1}\pi\right)\frac{w_x}{2 d_x}\right]^2 \times sinc\left[(\pi \tilde{f}_y y)\frac{w_y}{2 d_y}\right]^2$$

were used to fit the experimental data by use of a least squares fit. The term in square brackets together with the offset $c_0$ refers to $g^{(m)}$ for point-like sources. Thereby the amplitudes $A_i$ represent the relative strength of the corresponding spatial frequencies, what can be used to remove potential ambiguities in the image reconstruction if necessary. As real sources have a finite extension we multiply this term with an envelope function $env(x,y,w_x,w_y)$, which is an approximation to the exact analytic expression, where the envelope is registered by each fixed detector and is thus slightly shifted from each other[24]. However, the average of squared residuals between the approximation and the exact expression, which is far more complex to be fitted, is below 0.1 %. As in our experiment the six holes are identical, their extensions in the x- and y-directions can be deduced from the fit parameters $w_x, w_y$. The shift $\frac{m-2}{m-1}\pi$ in the x-component of the envelope results from the distribution of the fixed detectors at the MP along the x-axis. For correlations with the fixed detectors aligned at the MP along the y-axis, this shift transfers to the y-component of the envelope. Additionally there is a second envelope function $u_0 \times env(x, y, u_x, u_y)$, which approximates the change of $g^{(m)}$ due to different source intensities[24]. The fit parameters $u_0, u_x, u_y$ are needed for the fit, however, they are of no relevance for the image reconstruction.

From the fit curves we extracted a set of seven frequency vectors $\begin{pmatrix} f_{i,x} \\ f_{i,y} \end{pmatrix}$, representing $\zeta_{exp}$. Note that the two spatial frequency vectors $\begin{pmatrix} 1 \\ 1 \end{pmatrix}$ and $\begin{pmatrix} 1 \\ -1 \end{pmatrix}$ cannot be determined from $g^{(m)}(\mathbf{r}_1)$, if $m \geq 3$. Hence, there is a four-fold ambiguity to obtain the complete set of spatial frequencies $\zeta$ of the benzene structure from the set $\zeta_{exp}$, as either none, one of the two, or both spatial frequency vectors could be contained in $\zeta$. However, there is only one solution for completing this set, namely adding both frequency vectors $\begin{pmatrix} 1 \\ 1 \end{pmatrix}$ and $\begin{pmatrix} 1 \\ -1 \end{pmatrix}$, as none of the other three possibilities to complete $\zeta$ leads to a meaningful solution of the source arrangement in real space. The arrangement of the sources in real space is obtained by a computer algorithm that iteratively positions the sources and calculates the corresponding spatial frequencies until a source geometry is obtained that reproduces $\zeta$.

**Error discussion**

The statistical error of the experimentally obtained spatial frequencies was estimated by comparing measurements $g^{(m)}(\mathbf{r}_1; MP)$ of different correlation orders $m$ containing the same spatial frequencies. This leads to a statistical error of less than 1%. The systematical errors can be significantly larger, resulting mainly from the finite size of the CCD pixels (13.5 × 13.5 $\mu m^2$)[24]. Since the MP are assumed to be point-like, a finite pixel size results in an integration of the intensity pattern around the MP. Furthermore, the finite pixel size allows only for discrete values of the experimentally determined MP and consequently the lattice constants $d_x$ and $d_y$. For $d_x$ and $d_y$ (and thus $\tilde{f}_x$ and $\tilde{f}_y$) we derive a corresponding systematic uncertainty of 3.6% and 6.3%, respectively. Moreover, the width and height of the sources are extracted from the envelope function $env(x, y, w_x, w_y)$, which are governed by $d_x$ and $d_y$ as well. Both errors were taken into account in the image reconstruction for the positions of the individual sources (with respect to the center of the benzene structure) as well as the source extension (with respect to the respective source positions). In the final image of Fig. 3, the uncertainties are visualised by Gaussian color gradients in the x- and y-directions, with errors corresponding to the respective standard deviations. Since the origin of the image is chosen in the center of the benzene structure, a larger distance to the center leads to increasing errors in position, as lengths are determined in units of $d_x$ and $d_y$ and their respective uncertainties. Hence, the errors in position in y-direction for the two top and the two bottom sources are larger than for the left and right sources and vice-versa for the errors in position in the x-direction. The positional uncertainty is given by $(\Delta d_x, \Delta d_y) = (0.18\ \mu m, 0.53\ \mu m)$ for the two top and the two bottom sources and $(\Delta d_x, \Delta d_y) = (0.36\ \mu m, 0\ \mu m)$ for the left and right sources. Additionally the uncertainty for the extension of each source in x- and y-direction is $\Delta w_x = 0.11\ \mu m$ and $\Delta w_y = 0.20\ \mu m$, respectively.

**Explanation of the frequency filtering process**

To illustrate the frequency filtering process we consider the quantum mechanical expression of the m$^{\text{th}}$-order correlation functions $g^{(m)}(r_1, \ldots, r_m) \sim \langle \hat{E}^{(-)}(r_1) \ldots \hat{E}^{(-)}(r_m) \hat{E}^{(+)}(r_m) \ldots \hat{E}^{(+)}(r_1) \rangle$, where the positive frequency part of the field operator in the far field of N sources is given by $\hat{E}^{(+)}(r_j) \sim \sum_{i=1}^{N} e^{ik\,n_j \cdot R_i} \hat{a}_i$. Here, $n_j$ is the normalised vector pointing towards the detector at $r_j$, and $R_i$ and $\hat{a}_i$ are the position of source $i$ (on a lattice with lattice constants $(d_x, d_y)$) and the annihilation operator of a photon from source $i$, respectively. Due to the far field condition we can approximate the phase $k n_j \cdot R_i \approx 2\pi \tilde{f}_{i,x} x_j + 2\pi \tilde{f}_{i,y} y_j$. This leads to a separation in x- and y-components of the phase factors appearing in $g^{(m)}(r_1, \ldots, r_m)$. By placing the $m-1$ fixed detectors along the y-axis at $y_2 = \ldots = y_m = 0$, the y-dependent phase components vanish, reducing the further calculations to one dimension. With the $m-1$ fixed detectors placed along the x-axis at the MP, one can show that all phase terms vanish but those fulfilling the filtering condition $f_x = \kappa(m-1)\tilde{f}_x$, $\kappa \in \mathbb{N}_0$.[24] The y-components of the spatial frequencies are not accessed by the fixed detectors, but are included in the phase of the moving detector. Hence, although the filtering process targets only the x-components of the spatial frequencies, the y-components appear alongside with the x-components in the two-dimensional correlation function $g^{(m)}(r_1; MP_x)$. An analogue argumentation holds for $g^{(m)}(r_1; MP_y)$, i.e., placing the $m-1$ fixed detectors at the MP along the y-direction.

**Resolution power**

Our imaging algorithm requires a certain numerical aperture $\mathcal{A}$ in the detection plane to determine the spatial frequencies and thus to reconstruct the unknown emitter geometry. From classical optics and Abbes resolution limit we know that the numerical aperture required to image a source arrangement with spatial frequency set $\zeta$ is given by $\mathcal{A}_1^{(1)} \geq \frac{\lambda}{2 \cdot d_{min}}$, with $d_{min} = min_{f \in \zeta} \sqrt{|f_x \cdot d_x|^2 + |f_y \cdot d_y|^2}$, with $d_x$ and $d_y$ the lattice constants along the x- and y-direction, respectively. Obviously, $\mathcal{A}_1^{(1)}$ is determined by the smallest source-pair distance occurring in the source arrangement.

According to our algorithm, we resolve only a subset of the complete set of spatial frequencies $\zeta$ within a given correlation function $g^{(m)}(r_1; MP)$ of order $m$, that is we resolve either the smallest spatial frequency, associated with $d_{min}$, or larger frequencies. Therefore, the moving detector $D_1$ at $r_1$ will always require a numerical aperture $\mathcal{A}_1^{(m)} \leq \mathcal{A}_1^{(1)}$.

The numerical aperture $\mathcal{A}_{1,\ldots,m}^{(m)}$, required not only by the moving detector $D_1$ alone but by all detectors $D_1,\ldots, D_m$, including the m-1 detectors at the MP, depends on the grid constants $d_x$ and $d_y$. $\mathcal{A}_{1,\ldots,m}^{(m)}$ is smaller than $\mathcal{A}_1^{(1)}$ if the frequency vectors $\binom{0}{1}$ or $\binom{1}{0}$, corresponding to the smaller of the grid constants $d_x$ and $d_y$, is part of the complete set of spatial frequencies $\zeta^{24}$. Under these conditions our imaging algorithm provides a resolution below the canonical Abbe limit. If this condition is not met, a larger aperture is needed and our algorithm will not overcome the Abbe limit. In our particular arrangement of sources in the form of a benzene like hexagonal structure, the source geometry does not exhibit one of the spatial frequency vectors $\binom{0}{1}$ or $\binom{1}{0}$, so in this case we do not beat the resolution limit of classical optics.